\pdfoutput=1

\documentclass{aastex}
\usepackage{emulateapj5,epsfig}

\newenvironment{inlinefigure}{ 
\def\@captype{figure} 
\noindent\begin{minipage}{0.999\linewidth}\begin{center}} 
{\end{center}\end{minipage}\smallskip} 
\newcommand{\beq}{\begin{equation}}
\newcommand{\eeq}{\end{equation}}

\usepackage{graphicx}
\usepackage{amssymb}
\usepackage{epstopdf}

\font\tenbg=cmmib10 at 10pt

\def \rvecmu{{\hbox{\tenbg\char'026}}}
\def \rvecOmega {{\hbox {\tenbg\char'012}}} 

\def\lesssim{\mathrel{\hbox{\rlap{\hbox{\lower4pt\hbox{$\sim$}}}\hbox{$<$}}}}
\def\gtrsim{\mathrel{\hbox{\rlap{\hbox{\lower4pt\hbox{$\sim$}}}\hbox{$>$}}}}

\begin{document}
\title{Possible Signatures of Magnetospheric Accretion onto Young Giant Planets}

\author{R.V.E. Lovelace\altaffilmark{1}, K.R. Covey\altaffilmark{2,3,4}, and J.P. Lloyd\altaffilmark{2}}

\altaffiltext{1}{Departments of Astronomy and Applied and Engineering Physics,
Cornell University, Ithaca, NY 14853-6801;
RVL1@cornell.edu}

\altaffiltext{2} {Department of Astronomy,
Cornell University, Ithaca, NY 14853-6801;
kcovey@astro.cornell.edu; jpl@astro.cornell.edu}

\altaffiltext{3} {Hubble Fellow}
\altaffiltext{4} {Visiting Scholar, Department of Astronomy, Boston University, Boston MA 02215}

\begin{abstract}
Magnetospheric accretion is an important process for a wide range of astrophysical systems, and may play a role in the formation of gas giant planets.  Extending the formalism describing stellar magnetospheric accretion into the planetary regime, we demonstrate that magnetospheric processes may govern accretion onto young gas giants in the isolation phase of their development.   Planets in the isolation phase have cleared out large gaps in their surrounding circumstellar disks, and settled into a quasi-static equilibrium with radii only modestly larger than their final sizes (i.e., $ r \sim 1.4 r_{\rm final}$).  Magnetospheric accretion is less likely to play a role in a young gas giant's main accretion phase, when the planet's envelope is predicted to be much larger than the planet's Alfv\'en radius.  For a fiducial 1 M$_J$ gas giant planet with a remnant isolation phase accretion rate of $\dot{M}_{\odot} =$ 10$^{-10} M_{\odot}{\rm yr}^{-1}=10^{-7}M_{ J}{\rm yr}^{-1}$, the  disk accretion will be   truncated  at $\sim 2.7r_J$ (with $r_J$ is Jupiter's radius)
and drive the planet to rotate with a period of $\sim$7 hours.  
   Thermal emission from planetary magnetospheric accretion will be difficult to observe; the most promising observational signatures may be non-thermal, such as gyrosynchrotron radiation that is clearly modulated at a period much shorter than the rotation period of the host star.
\end{abstract}

\keywords{accretion, accretion disks --- planets ---etc}

\section{Introduction}

Magnetically controlled accretion (e.g. Ghosh \& Lamb 1979) governs mass accretion onto objects ranging from white dwarfs and neutron stars (Warner 2004) to supermassive black holes (Koide et al. 1999).  Magnetospheric accretion models (K\"onigl 1991) explain many of the observational characteristics of young stars, such as the presence and magnitude of UV excesses from accretion shocks (Calvet \& Gullbring 1998; Herczeg \& Hillenbrand 2008) and the kinematic structure and rotational modulation of spectral line profiles arising from accretion funnel flows (e.g. Bouvier et al. 2003, Kurosawa et al. 2006).

Hydrodynamical simulations of mass accretion onto young giant planets have been performed (e.g., Papaloizou \& Nelson 2005; Hubickyj, Bodenheimer, \& Lissauer 2005;  Ayliffe \& Bate 2009), but these analyses have not formally considered the influence of the planet's magnetic field.  Young gas giants likely possess strong dynamos, driven by rapid rotation and significant convective motions in their interiors (S\'anchez-Lavega 2004).  The appreciable large-scale magnetic fields produced by these dynamos could channel subsequent accretion onto the giant planet, potentially producing observational signatures analogous to those observed in young stars. 

   Earlier, Quillen and Trilling (1998) and Fendt (2003) discussed 
magneto-centrifugally driven outflows from  the circumplanetary disks.  Fendt considered planetary accretion rates $\sim 10^{-6}M_\odot{\rm yr}^{-1}
= 10^{-3}M_J{\rm yr}^{-1}$.     Our analytic calculations, combined with recent estimates of proto-planetary radii, imply that the planet's magnetic field is likely to be buried for such a high accretion rate.  
That is, the planet's physical
radius will be larger than the magnetic standoff or Alfv\'en
radius.
     The focus of the present work is on the conditions  with
much slower accretion ($\sim 10^{-10}M_\odot {\rm yr}^{-1}$),
where the planet's magnetic field is not buried so that it
may  influence  the planet's accretion.   This value
of the accretion rate agrees with that considered by Quillen and Trilling (1998). 
   The present work is concerned with the possible time-dependence
of the rotating planet's emission due to its non-symmetric magnetosphere
rather than its outflows.

Our interest in this problem was motivated by recent observations of a young stellar object, IC1396A-47, whose mid-IR light curve reveals periodicities on two very different timescales (Morales-Calder\'on et al. 2009).     The long-period variability ($P_{\rm rot} \approx 9$ d;  $\delta$[3.6] $\sim$ 0.2 mag.) is likely due to the rotation of a hot spot on the surface of the star, but the origin of the short-period component ($P_{\rm rot} \approx$ 3.5 hr.; $\delta$[4.5] $\sim$ 0.04 mag.) is less clear.  Morales-Calder\'on et al. conclude that $\delta$ Scuti pulsations are the most likely explanation for the short--period variability, but the timescale is also within the range of rotation periods expected for young planets.  

In this letter, we explore potential observational signatures of magnetospherically accreting young planets.  In Section 2 we derive analytic expressions that describe magnetospheric accretion in the planetary regime.  With few observational constraints on the physical properties of young gas giants, we investigate in Section 3 predictions of our magnetospheric accretion model over the area of parameter space potentially inhabited by proto-gas giants.  We explore potential observational signatures of this process in Section 4, and summarize our conclusions in Section 5.  

\section{Magnetospheric Accretion to Giant Planets}

       We consider the early stage ($\lesssim 10$ Myr) of a giant planet of mass $M_p$ (of the order of Jupiter's mass) orbiting a young star surrounded by an accretion disk.  Giant planets are thought to form at $R_p \sim 10-25$ AU from the star and then undergo Type II inward migration at roughly the viscous accretion speed of the disk (Papaloizou et al. 2007). 
       
             Critical parameters for magnetospheric accretion by a 
 giant planet are the planet's characteristic radius, where an accretion shock may form, and its accretion rate.
     These parameters are strongly  time-dependent during the planet's
formation.   
  Planets with high mass accretion rates ($\sim 10^{-8}
M_\odot{\rm yr}^{-1}$) are predicted to have radii much larger than Jupiter's radius
(denoted $r_J$; Hubickyj, Bodenheimer, \& Lissauer 2005; Helled \& Schubert 2008; Ayliffe \& Bate 2009).    The planet quickly cools once the mass accretion slows significantly, however, with models predicting a rapid collapse to much smaller radii ($\sim r_J$; Lissauer et al. 2009). 
      
      In the core accretion model, a gas giant grows as its rocky core accretes material within its Hill radius, $r_H=R_p[M_p/(3M_*)]^{1/3}$, where $M_*$ is the star's  mass, and $M_p$ is the planet's mass (e.g., Hubickyj, Bodenheimer, \& Lissauer 2005).  The planet's Hill radius expands as its mass increases, enabling the planet to accrete mass from a larger portion of the circumstellar disk.  The planet's mass accretion rate thus increases until its Hill radius exceeds the scale height of the circumstellar disk.  This isolates the planet within a gap in the circumstellar disk, at which point Type II migration begins (e.g., Papaloizou et al. 2007).  This  is sketched in Figure 1.

\begin{inlinefigure}
\centerline{\epsfig{file=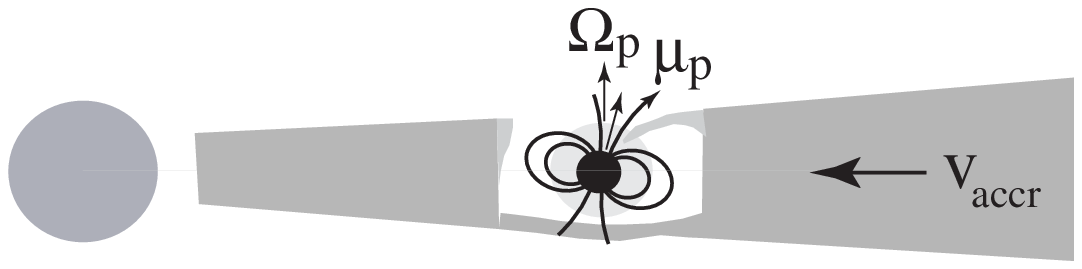,height=1.3in,width=3.in}}
\epsscale{0.8}
\figcaption{Envisioned geometry of a giant, gap-forming planet
in the accretion disk of a young star.}
\end{inlinefigure}

The gap the planet opens in the circumstellar disk
does not stop accretion to the planet (e.g.,
Alexander \& Armitage 2009).  Tidal streams  flow across the gap in the disk, sustaining the mass accretion rate onto the planet ($\dot{M}_p$) and the star ($\dot{M}_*$) at some fraction of the disk accretion rate expected in the absence of a planet ($\dot{M}_{\rm d}$; Lubow et al 1999, D'Angelo et al. 2002).  

      When the planet's radius $r_p \ll r_H$,
matter accreting to the planet will in-spiral in a Keplerian disk from roughly $r_H/3$ to the planet's surface (Ayliffe \& Bate 2009).   The nominal {\it accretion luminosity} of this circumplanetary disk is
\begin{equation}
L_{pd} ={GM_p\dot{M}_p\over r_p }
 \approx  10^{29}~
 {1\over{\cal R}_p}{M_p\over M_J}{\dot{M}_p\over \dot{M}_0}~
 {{\rm erg}\over {\rm s}}~.
 \end{equation}
Here, $M_J$ is Jupiter's mass ($1.9\times 10^{30}$ g),  $r_J$ is Jupiter's radius ($7.14\times 10^{9}$ cm) and
${\cal R}_p \equiv r_p/r_J$.   We
assume ${\cal R}_p \approx 1.3$, as found by 
Fortney, Baraffe, \& Militzer (2009) at $t=10^7$yr.

\subsection{Magnetic disk locking of planets }

     If the planet is unmagnetized and accretes $\gtrsim$10\%
of its mass from  a  circumplanetary disk, it will spin-up
to a  break-up angular velocity 
$\Omega_{\rm max}=    (GM_p/r_p^3)^{1/2}$.
This break-up velocity corresponds to a minimum rotation period
of $T_{\min} \approx 3(M_J/M_p)^{1/2}{\cal  R}_p^{3/2}{\rm hr}\approx
4.4$hr for $M_p=M_J$ and ${\cal R}_p=1.3$.
   Accretion to a planet rotating at break-up continues via mechanisms describing of disk accretion to  a rapidly rotating 
unmagnetized   star (Bisnovatyi-Kogan 1993).
     
   Magnetic processes may influence accretion onto the planet.
   The planet's rapid rotation and convective interior suggest strong dynamo activity that could produce appreciable large-scale, dipole magnetic fields (S\'anchez-Lavega 2004). 
        As a reference value we adopt a magnetic moment for young gas giants of $10$ times the
magnetic moment of Jupiter ($\mu_J$; S\'anchez-Lavega 2004), or a surface magnetic field of about
$85$G at the magnetic pole for a Jupiter size planet.
    This magnetic field may be strong enough to truncate the circumplanetary disk at the Alfv\'en radius, where the kinetic energy density
of the disk plasma ($\rho v_K^2/2,$ with $v_K=\sqrt{GM_p/r_p}$, 
the disk's Keplerian velocity)
is equal to the energy density of the magnetic field
(${\bf B}^2/8\pi$; Davidson \& Ostriker 1973; Elsner \& Lamb 1977;
Long, Romanova, \& Lovelace 2005).   The Alfv\'en radius of an accreting, {\it non-rotating planet} ($r_{A0}$) can be expressed as

\begin{eqnarray}
r_{A0} &\approx& \left({\mu_p^2 \over \dot{M}_p \sqrt{G M_p}}\right)^{2/7}
\nonumber\\
&\approx& 10^{10}\left({\mu_p \over 10 \mu_J}\right)^{4/7}
\left({\dot{M}_0 \over \dot{M}_p}\right)^{2/7}\left({M_J \over M_p}\right)^{1/7}
\!\!\!\!{\rm cm}
\end{eqnarray}
where $\mu_J$ is the magnetic moment of Jupiter and $\mu_p$ is the magnetic moment of the young gas giant.

    A rotating giant planet's effective Alfv\'en radius is actually somewhat larger than that implied by the first order calculation in Equation 2. This effect can be understood by noting that the Alfv\'en radius is the distance where the kinetic energy of  the disk plasma in the {\it reference frame rotating with the planet} is equal to the magnetic energy density.  This implies that the effective Alfv\'en radius will be larger for rapidly rotating planets, as the planet's rotation will reduce the velocity contrast between the planet's magnetic field and the  Keplerian motion in the circumplanetary disk.  The effective Alfv\'en radius for a rotating planet can be calculated  as
$r_A = r_{A0}[1-\Omega_p r_A/v_K(r_A)]^{-4/7}$ or 
$ \eta_A^{-7/2}= \big(1-g\omega_p \eta_A^{3/2}\big)^2$,
where $\eta_A\equiv r_A/r_{A0}$, 
$\omega_p\equiv \Omega_p/\Omega_{\rm max}$ with
$0\leq \omega_p \leq 1$, and
 $g \equiv \Omega_{\rm max} (GM_p/r_{A0}^3)^{-1/2}$.  

    The relation between $\eta_A$ and $\omega_p$
can be rewritten as 
\begin{equation}
\omega_p =
(\eta_A^{-3/2} \pm \eta_A^{-13/4})g^{-1}~.
\end{equation}
   The solid lines in Figure 2 show the relationship between $\omega_p$ and
$\eta_A$ for the reference values of equation (2), which
give $g=0.636$.    We do not consider tidal locking of the planet's rotation, which is estimated
to take longer than $10$Myr for $R_p >0.1$ AU.

\begin{inlinefigure}
\centerline{\epsfig{file=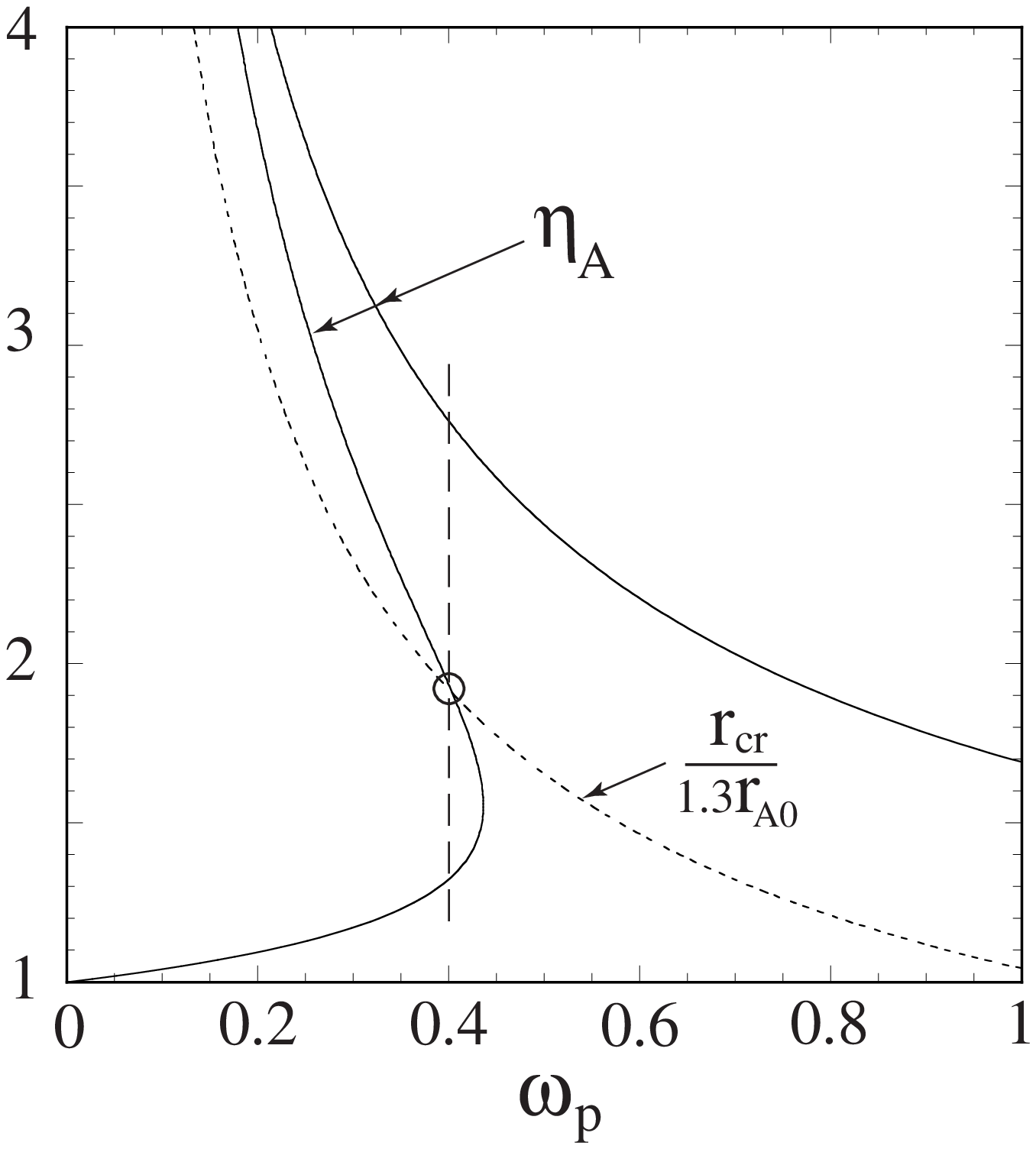,
height=2.5in,width=2.5in}}
\epsscale{0.8}
\figcaption{The solid curves
show the relation between $\eta_A=r_A(\omega_p)/r_{A0}$ and 
$\omega_p = \Omega_p/\Omega_{\rm max}$  from
equation (3).    
   The dotted curve shows $r_{\rm cr}(\omega_p)/(1.3 r_{A0})$.
The intersection of the dotted and solid curve corresponds to
disk locking as discussed in the text.}
\end{inlinefigure}

While the planet's rotation influences where the circumplanetary disk is truncated, transfer of angular momentum between the planet and circumplanetary disk via magnetic interactions also influences the planet's rotation rate. 
This feedback mechanism drives towards an equilibrium state where there is no net angular momentum flow between the planet and the inner edge of the circumplanetary disk (Ghosh \& Lamb 1979).  Simulations of magnetospheric accretion have identified that stars in this `disk-locked' state rotate more slowly than break-up ($\Omega _p< \Omega_{\rm max}$), such that the Alfv\'en radius slightly exceeds the `co-rotation radius' ($r_{\rm cr}$), where the angular velocity of Keplerian motion in the disk equals the star's angular velocity (i.e., $r_{\rm cr}\equiv (GM_p/\Omega_p^2)^{1/3} \approx 1.6\times 10^{10} (M_p/M_J)^{1/3} (\Omega_J/\Omega_p)^{2/3} {\rm cm} \approx 1.3 r_A$; Long et al. 2005).  This is equivalent to  the condition that $r_{\rm cr}(\omega_p)/(1.3 r_{A0})=
\eta_A(\omega_p)$.  
    Figure 2 shows both $r_{\rm cr}(\omega_p)/(1.3 r_{A0})$ (dotted
curve) and $\eta_A(\omega_p)$ (solid curves from equation 3).    
  Their intersection point (indicated by the circle)  corresponds to disk-locking. 
   Along the vertical dashed line {\it above} the circle,
where the line intersects at $\eta_A > r_{\rm cr}/(1.3 r_{A0})$, the planet will {\it spin-down},
whereas below the circle where the line intersects
at $\eta_A < r_{\rm cr}/(1.3 r_{A0})$, the planet will {\it spin-up}
(see Lovelace, Romanova, \& Bisnovatyi-Kogan 1999).

\subsection{Magnetospheric accretion from the circumplanetary disk}

      The matter inflow from the inner edge of the circumplanetary disk at $r_A$ to the
 planet's surface at $r_p$ occurs in narrow ``funnel streams''  
which follow the planet's magnetic field as
proposed by Ghosh and Lamb (1979) and observed
and analyzed in  three-dimensional magnetohydrodynamic (MHD)
simulations by Romanova et al. (2002, 2004).    
    If the planet's magnetic dipole moment $\rvecmu_p$ is
not greatly misaligned with the planet's rotation 
axis $\rvecOmega_p$ (assumed perpendicular to the disc around the
star), then the funnel streams  will impact the planet's
surface close to the magnetic poles 
roughly perpendicular to the surface.   
   The funnel streams are approximately stationary
in a reference frame rotating with the planet
(Romanova et al. 2004).   
  Thus, the flow speed along the funnel
stream $u_f(r)$ is obtained from
Bernoulli's law in a reference frame
rotating with angular rate $\Omega_p$.
This gives 
\begin{equation}
{u_f^2\over 2} + \Phi(r_p) ={1\over 2}\big[v_K(r_A)- \Omega_pr_A\big]^2
+\Phi(r_A)~, 
\end{equation}
where $v_K(r_A)$ is the
azimuthal velocity of the disk matter at $r_A$ and
$\Phi(r)=-GM_p/r -\Omega_p^2 r^2/2$ is the effective
potential.  

   The impact of a funnel stream  on the planet's
surface will create a strong shock wave
covering a small fraction of the planet's surface -
a  ``hot-spot'' -
where the power in the stream is thermalized
and radiated away.   
  Just outside the shock wave we have $\dot{M}_p =
2\pi r_{fp}^2 \rho_{fp} u_{fp}$, where $r_{fp}$
is the radius of the funnel stream at the planet's
surface and $\rho_{fp}$ is the density before
the shock.  Translating into the planetary regime the
 results of the 3D MHD simulations of stellar mass accretion described above 
implies that $r_{fp} = (0.1-0.2) r_p$ (Romanova et al. 2004).
    The power dissipated behind the shock is
assumed to be radiated as black-body radiation
so that the effective temperature of the hot-spot
is simply
\begin{equation}
T_{\rm eff} =\left({\rho_{fp}u_{fp}^3 \over
2 \sigma}\right)^{1/4}~,
\end{equation}    
where $\sigma =5.67\times 10^{-5}$ cgs is the Stefan-Boltzmann
constant.   

      Figure 3 shows an example of a funnel flow obtained from
three dimensional magnetohydrodynamic simulations by
Romanova, Kulakrni, \& Lovelace (2008).
  
\begin{inlinefigure}
\centerline{\epsfig{file=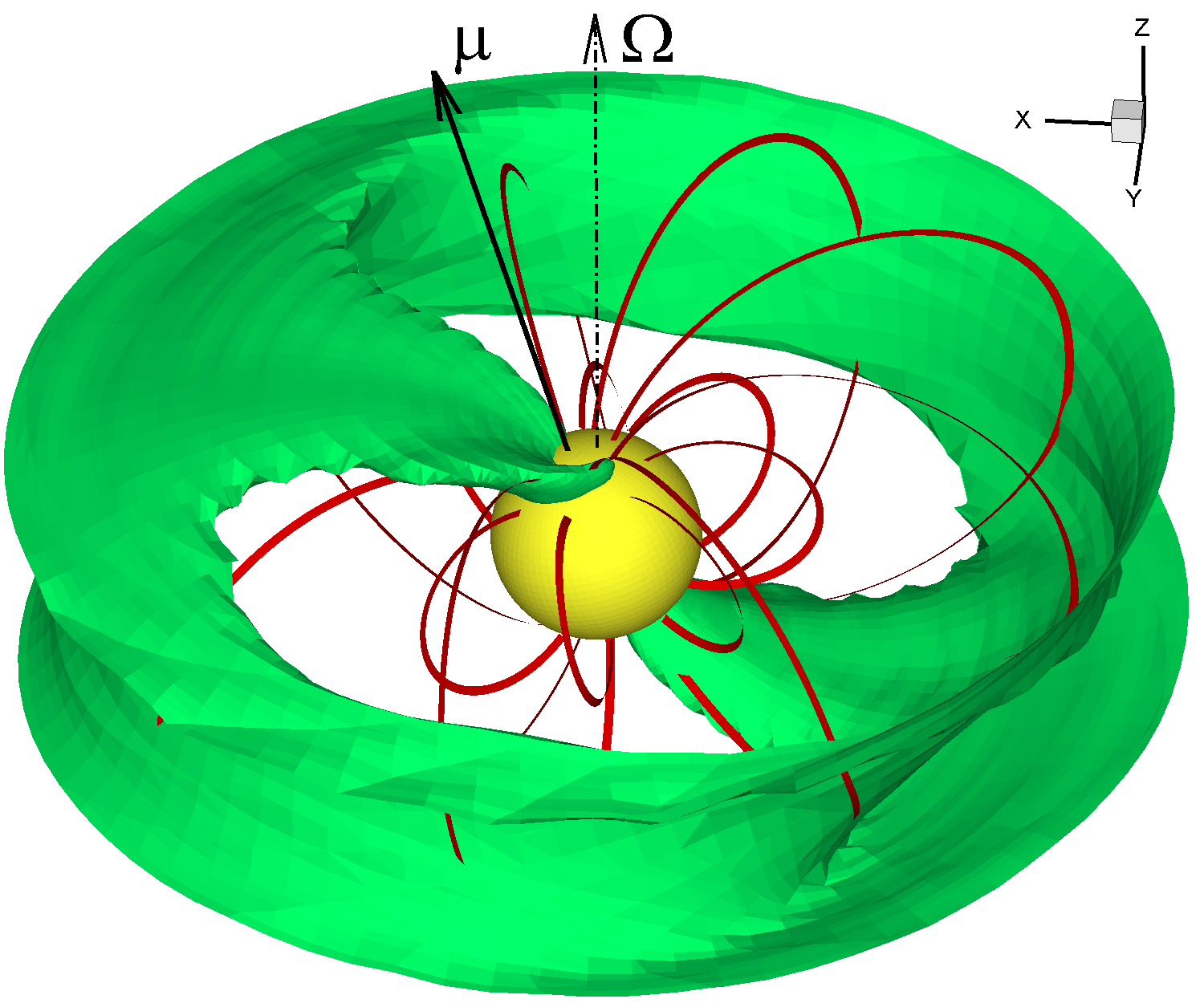,
height=3.in,width=3.in}}
\epsscale{0.8}
\figcaption{Funnel flow onto a
rotating magnetized object from 3D MHD simulations by Romanova,
Kulkarni, and Lovelace (2008).  The red lines are magnetic field
lines,  $\rvecmu$ is the magnetic moment, $\rvecOmega$ is the
 angular velocity vector.  The green color labels an isodensity
 surface.  Inside of the green surface the density is higher.}
\end{inlinefigure}

\subsection{Parameters of Accreting Magnetized Giant Planets}

      The previous subsections give a general 
description of   accreting, rotating magnetized giant planets.  
    The focus of the present work is on the conditions
where the magnetic field of the rotating planet influences
the planet's accretion. 
   Evidently, the planet's magnetic field can have a significant
influence on its accretion only under conditions where
the planet's Alfv\'en radius exceeds its physical radius.
Otherwise, the magnetic field will be too weak to channel the accreting material into funnel streams. 
 As is clear by inspection of Equation 2, the most 
 time-variable parameter governing a planet's Alfv\'en radius is  its accretion rate, which is likely to change more rapidly than its mass or magnetic moment.  
     To explore the area of parameter space in which magnetic effects will be important, we plot in Figure 4 the Alfv\'en radius and possible planet radii
 (discussed below)
 as a function of the planet's accretion rate
 $\dot{M}_p$ measured in units of $\dot{M}_0 \equiv 10^{-10}M_\odot
 {\rm yr}^{-1}$.   For this plot we assume
 $M_p=M_J$ and $\mu_p = 10 \mu_J$.

  Theoretical models of gas giant protoplanets have been developed by
Papaloizou and Nelson (2005).  
 They consider two models, an early stage of planet
formation - termed type A - where the planet is an extended structure going
out to its Hill sphere,  and a later stage - termed type B - where the
the protoplanet is much more compact with a free surface which accretes
from a circumplanetary disk.   Both models start from solid cores of
$5$ or $15M_\oplus$.   The type A models are found to have long
mass accretion time-scales for $M_p  \lesssim 30 M_\oplus$ and
they may undergo rapid type I inward migration unless this migration
is suppressed (e.g., Rafikov 2002; Li et al. 2009).
    In the type B models the planet's radius 
approaches $\sim 2 \times 10^{10}$ cm and is relatively independent
of the accretion rate $\dot{M}_p$ which is determined by star's accretion disk.   The planet undergoes slower type II migration.
    In Figure 4 the horizontal dashed lines indicate the planet's radii for type
A and B models.

Numerical simulations of planet formation also give information
on the physical radii of young gas giants as functions of mass and time.  
Hubickyj et al. (2005) and Ayliffe \& Bate (2009) find that a young ($t<2$ Myrs) proto-Jupiter ($M_p= M_J$) has a typical mass accretion rate of $\sim 3 \times 10^{-8} M_{\odot}$/yr, and a  radius of $\sim 100 R_J$ which is of the order of the radii of the type A models of Papaloizou and Nelson (2005). 
  It is not clear why the simulations do not give the Papaloizou and Nelson 
type B solutions.
     Figure 4 includes the line marked C for 
an older  ($t > 10$ Myr)  non-accreting Jupiter-mass
planet  predicted by Fortney et al. (2009) to have a radius $\sim 1.3 r_J$.

From Figure 4, it is apparent that planets with accretion rates $\gtrsim
M_0 \equiv 10^{-10}{\rm M}_{\odot}$/yr have physical radii larger than their Alfv\'en radii so that the planet's accretion will not be affected by its magnetic field.
   Magnetospheric accretion may, however, be important once a gas giant ends its main accretion phase (type A models  of Papaloizou \& Nelson, 2005), having entered a type B model with an extended circumplanetary disk.   For the accretion rates considered by Fendt (2003)
   ($\sim 10^{-6} M_\odot{\rm yr}^{-1}$ ), we expect that the planet's
radius is larger than the Alfv\'en radius.

      In the following we consider a fiducial model with
$\dot{M}_p =\dot{M}_0  \equiv 10^{-10}M_\odot{\rm yr}^{-1}
\approx 10^{-7}M_J~{\rm yr}^{-1}$.
    In the equilibrium disk-locked state,  the planet
rotates with $\omega_p \approx 0.4$ and truncates its circumplanetary disk at $r_A =1.9 r_{A0}$.  
      For our fiducial gas giant planet, this corresponds to a rotation period  of $2\pi/\Omega_p \approx 7.4$ hr, somewhat less than Jupiter's  rotation period of $9.8$ hr.
For our adopted reference values this gives a flow
speed at the planet's surface
$u_{f}(r_p)=u_{fp}\approx 21$ km s$^{-1}$.   

\begin{inlinefigure}
\centerline{\epsfig{file=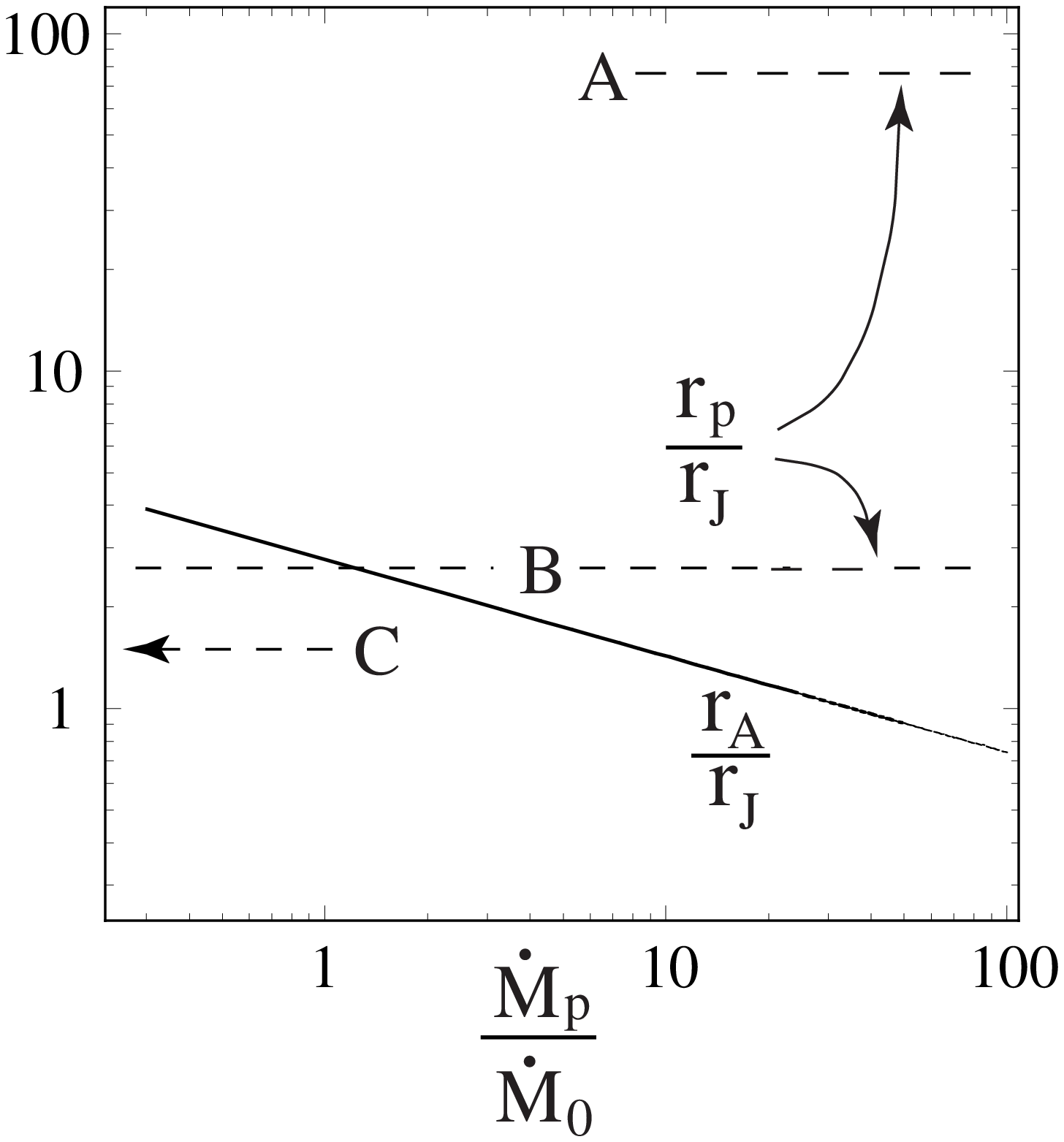,height=2.5in,width=2.5in}}
\epsscale{0.8}
\figcaption{The solid curve gives
the effective Alfv\'en radius $r_A \approx 1.9 r_{A0}$ in units
of Jupiter's radius $r_J$  from Eq. 2 
assuming the planet's magnetic moment is
$10\mu_J$. The factor of $1.9$ is discussed
in the text.    Here, $\dot{M}_0 =10^{-10}M_\odot/{\rm yr} \approx
10^{-7}M_J/{\rm yr}$.
The dashed curve shows a sketch of
the planet's radius $r_p$ in units of $r_J$ estimated from
the work of Hubickyj et al. (2005)  for times $t \lesssim 2 {\rm Myr}$
and from the work of Fortney et al. (2009) for $t\gtrsim 10{\rm Myr}$.}
\end{inlinefigure}

\section{Observational Signatures of Accreting Giant Planets}

We now use the analytic framework described 
above to consider potential observational signatures 
of planetary magnetospheric accretion.  In turn, we explore the basic thermal 
signature of this process, the potential to spatially or temporally resolve 
this thermal signature, and finally non-thermal observables. 

{\bf Thermal Signatures:}
The equations in section 2 allow us to describe the basic thermal properties of 
the accretion shock and circumplanetary disk associated with a
giant planet undergoing magnetospheric accretion.  
The accretion shock is likely the warmer of the two components, and thus
more amenable to observations at the shorter wavelengths accessible
to ground based observatories.  For our reference values and $r_{fp}=0.2 r_p$ we
find the accretion shock's $T_{\rm eff}\approx 2100$ K, whereas for
$r_{fp}=0.1 r_p$, $T_{\rm eff} = 3000$ K.   This temperature is somewhat less 
than the temperature of a typical T Tauri star photosphere ($T_{\rm eff}\sim4000K$; 
Johns-Krull \& Gafford 2002), providing some leverage for decomposing 
these two components of a star's spectral energy distribution.  

The true challenge of detecting the thermal signature of a planetary accretion shock
becomes apparent, however, when one considers the overall luminosity of the shock in 
contrast to that of the stellar photosphere.  The power dissipated in one hemisphere's 
accretion hot-spot is  $\dot{M}_p u_{fp}^2/4$
which is about $0.063 L_{pd}$ for our fiducial model.  As  
$L_{pd}$ is on the order of 10$^{29}~ {{\rm erg}/ {\rm s}}$,
this indicates an observable accretion luminosity of $\approx10^{27-28}~ {{\rm erg}/{\rm s}}$.  T Tauri stars, however, have typical photospheric luminosities of $L_* \approx 10^{33}$ erg s$^{-1}$, and young giant planets are thought to have luminosities of 10$^{27-30} {\rm erg s^{-1}}$ (Hubickyj et al. 2005; Marley et al. 2007).  Detecting the thermal signature of a planetary accretion shock will therefore require disentangling three blackbody spectra to identify a component that contributes just one of every million photons detected from a given young stellar object.

The other thermal signature of planetary magnetospheric accretion will arise from the circumplanetary disk.  The effective temperature of the disk, 
which we assume to be optically thick, is 
$T_{\rm effp}=[3GM_p\dot{M}_p/(16\pi \sigma)]^{1/4}
\approx 1200(r_J/r)^{3/4}(M_p/M_J)^{1/4}
(\dot{M}_p/\dot{M}_0)^{1/4}$ K, where $r$ is
the distance from the planet's center.
     If the circumplanetary disk is truncated at $r_A \sim 2 r_{A0} \sim 2.7 r_J$, the inner edge of the disk is predicted to have a temperature of $\sim 1200$K , much higher than than the local temperature
of the star's disk for $R_p \gg 0.25$ AU.   
    The ratio of the luminosity of the planet's disk
($GM_p\dot{M}_p/2r_p$) to the luminosity of
the annular region of the star's disk of with $2r_H$
(in the absence of the planet) is
$(M_p/3M_*)(\dot{M}_p/\dot{M}_*)(R_p/r_p) (R_p/r_H)$.
     For a giant planet of Jupiter's mass and semimajor
 radius and $M_*=M_\odot$, this ratio is 
 $\sim  50(\dot{M}_p/\dot{M}_*)$.  While the circumplanetary disk may be more luminous than the surrounding material in the circumstellar disk, the emission from the circumplanetary disk may be difficult to disentangle from the young planet itself, which could well have a similar temperature and luminosity ($L\sim 10^{-6}$ to $ 10^{-3} L_{\odot}$, $T_{\rm eff} \sim 500~{\rm to}~2000$K; Marley et al. 2007).  Spatially resolving a truncated circumplanetary disk would provide a clear observable signal, but resolving spatial scales of $ r_J$ in Taurus ($d\sim140$ pc; Loinard et al. 2007) requires high contrast observations on angular scales of 10${-6}$\arcsec, well beyond current observational capabilities. 

     The hot-spots can give periodic variations in the
observed radiation depending on the    
angle $\Theta$ between the planet's
magnetic moment $\rvecmu$ and its rotation axis
$\rvecOmega_p$ {\it and} the angle $\iota$ between
the line-of-sight to the object and $\rvecOmega_p$
as analyzed by Romanova et al. (2004).  
    This work assumes that the planet's rotation axis
is the same as the rotation axes of the disk and the star.
   Thus we do not consider close-in  giant planets where
the planet's  orbital  angular momentum
is in some cases observed to be misaligned with the
star's angular momentum (e.g., Triaud et al. 2010).
   For $\Theta=0$ the planet is axisymmetric
and there are no variations in the luminosity for any
value of $\iota$.
   For $\Theta \neq 0$ and $\iota =0$ the disk is face-on and there 
are no luminosity variations.      
     For $\Theta \neq 0$ and $\iota$ larger than a critical
value (possibly $\sim 45^\circ$) dependent on the geometry of the 
gap in the disk the radiation from the hot spot will be obscured
by the disk.
    Note that the misalignment or tilt angle of Jupiter's magnetic
field is $\Theta \approx 10^\circ$.  

To  estimate of the magnitude of the fluctuations
from a rotating giant planet we assume
the host star to be a typical K7/M0 Classical T Tauri Star, and for specificity adopt the physical parameters measured for AA Tau: $R_* = 1.67 R_{\odot}$, $T_{\rm eff} = 4000$ K, and $P_{\rm rot}$=8 days (Johns-Krull \& Gafford 2002).   Thus the star's photospheric luminosity is
$L_* \approx 2.5\times 10^{33}$ erg s$^{-1}$.
   To get an upper limit on the detectability of accretion onto a giant planet, we first assume $100\%$  rotational modulation of the emission
from the planet's `hot spot'.
      For one hemisphere of
the planet this is $L_{hs+} \lesssim GM_p\dot{M}_p/(4r_p)$,
where $M_p$, $\dot{M}_p$, and $r_p$ are the
planet's mass, accretion rate, and radius.   
    The rotation period of the planet ($3-10$ hr) is
not affected by the tidal interactions in $10$ Myr 
for major radii $> 0.1$ AU.    
   We neglect the radiation from the star's
accretion disk and accretion shock  and the planet's disk.   
    The fractional variation in the flux due to the 
rotating planet is
\begin{equation}
{ L_{hs+} \over L_*} \lesssim
    10^{-5} \left({ M_p \over M_J}\right)
    \left({\dot{M}_p \over 10^{-10} M_\odot {\rm yr}^{-1}}\right)
    \left({r_J \over r_p}\right)~,
\end{equation} 
such that detecting this effect will require exquisitely precise, stable photometry.

 {\bf Non-Thermal Signatures:}
   For $r_A >r_p$, note
that an accreting magnetized planet may 
produce oppositely
directed  conical winds and jets along its rotation 
axis $\rvecOmega_p$.   
    Analogous winds  and jets are common
features of young accreting magnetized stars (see
review by Ray et al. (2007), and they have been
studied in 2D and 3D magnetohydrodynamic
simulations (e.g.  Romanova et al. 2009).   
In agreement with estimates by
Quillenn and Trilling (1998) and Fendt (2003), we expect the mass outflow rate
to be about $0.1 \dot{M}_p$ and the outflow
velocity to be $\sim v_K(r_p)$ which is about
$42$ km/s for a Jupiter mass planet and significantly
larger than the escape speed from the star.
      It is also possible that the giant planet has
magnetic reconnection flares analogous to
those observed in young stars (e.g.,
Getman et al. 2008).      
     However,  the magnetic
field strength of the planet is down by more than a factor
of $10$ from that of the
star and the emission volume ($V_e \propto R_p^3$)
is down by a factor of $10^3$ so that the flare
energy ($\propto B^2 V_e$) is down by more than a factor
of $10^5$ compared with that of the star.
   On the other hand high energy electrons
produced in a flare are expected to  propagate 
both outward and downward along the planet's
field lines.  The electrons reaching
 the planet's magnetic poles 
can give rise to gyrosynchrotron radiation in
the radio band ($\sim $GHz).
  This type of circularly polarized radiation has
been observed from T Tau S 
(Skinner \& Brown 1994) and from embedded young class I
objects (Feigelson, Carkner, \& Wilking 1998;
Choi et al. 2009).   The radiation is thought to
be analogous to that from RS CVn binary systems
(e.g., Morrris, Mutel, \& Su 1990).
   Because this radiation (unlike the X-ray)
comes from near the planet's surface with
the emission $\propto B^2$, it may be
modulated by the planet's rotation depending
of course on $\Theta$ and $\iota$.

\section{Conclusions}
In this paper, we provide a first extension of the formalism describing stellar magnetospheric accretion into the planetary regime.  In doing so, we demonstrate that:\\
1. Magnetospheric processes may govern accretion onto young gas giants in the isolation phase of their development. Magnetospheric accretion is less important for younger gas giants in their main accretion phase: thermal energy from accretion inflates their envelopes to sizes much larger than their magnetic Alfv\'en radius.\\ 
2. For a fiducial 1 M$_J$ gas giant planet with a remnant isolation phase accretion rate of $\dot{M}_{\odot} =$ 10$^{-10}$ M$_{\odot}$, magnetospheric accretion will truncate the circumplanetary disk at $\sim$2 Alfv\'en radii (corresponding to $\sim$1.5 plantary radii and $\sim$3 Jupiter radii), and drive the planet to rotate with a period of $\sim$7 hours.  \\
3. Thermal emission from planetary magnetospheric accretion will be difficult to observe due to the stringent flux contrast and/or spatial resolution required for such measurements. The most promising observational signatures may be non-thermal, such as gyrosynchrotron radiation that is clearly modulated at a period much shorter than the rotation period of the host star.

\section*{Acknowledgements}

We thank M.M. Romanova, P.D.
Nicholson, Sally Dodson--Robinson, M.S. Tiscareno,  M.M. Hedman,
and E.D. Feigelson
 for valuable comments and discussions.
RVEL  was supported in 
part by NASA grant NNX08AH25G and by
NSF grants AST-0607135 and AST-0807129.
Support for this work was provided by NASA through Hubble Fellowship grant HST-HF-51253.01-A  awarded by the Space Telescope Science Institute, which is operated by the Association of Universities for Research in Astronomy, Inc., for NASA, under contract NAS 5-26555.

\end{document}